\documentclass{JINST}
\usepackage{graphicx,natbib,amssymb,lineno, epsf}

\title {Characterization of large area photomultipliers 
and its application to dark matter search with noble liquid detectors}

\author{A.~Bueno, J.~Lozano, A.J.~Melgarejo, F.J.~Mu\~noz,
J.L.~Navarro, S.~Navas and A.G.~Ruiz \\
Dpto. de F{\'i}sica Te\'orica y del Cosmos \& C.A.F.P.E.,
Universidad de Granada, 18071 Granada, Spain\\}

\abstract
{
There is growing interest in the use of noble liquid 
detectors to study particle properties and search for 
new phenomena. In particular, they are extremely suitable 
for performing direct searches for dark matter. In this kind 
of experiments, the light produced after an interaction 
within the sensitive volume
is usually read-out by photomultipliers. The need to go to masses 
in the tonne scale to explore deeper regions of the parameter 
space, calls for the use of large area photomultipliers. 
In this paper we address the need to perform laboratory
calibration measurements of these large photomultipliers, 
in particular to characterize its behaviour at cryogenic 
temperatures where no reference from the manufacturer is available.
We present comparative tests of phototubes
from two companies. The tests are performed in conditions similar to
those of operation in a real experiment.
Measurements of the most relevant phototube parameters 
(quantum efficiency, gain, linearity, etc.) 
both at room and liquid Argon temperatures are reported.
The results show that the studied phototubes comply with the stringent 
requirements posed by current dark matter searches 
performed with noble-liquid detectors.
}


\begin{document}

% ==================================
\section{Introduction}
\label{sec:intro}
% ==================================

Particle interactions in liquefied noble gases produce charge by means of
{\it ionization} of the atoms of the medium and {\it light}, by the de-excitation
of the formed dimers~\cite{Doke1}. This makes this kind of detectors very 
suitable tools to search for new physics. In particular, they represent a competitive
technology in the quest for dark matter~\cite{Gaitskell:2004}, since in 
this kind of experiments it is crucial to have several independent 
observables in order to suppress background events. 

There are several ways to collect both charge and
light~\cite{Bellazzini:2007} in noble liquid detectors. 
For light readout, the use of photomultiplier tubes (PMTs)
is highly recommended as their time resolution allows to distinguish photons
produced in a triplet state dimer (slow component) from those
produced in a singlet state (fast component)~\cite{Kubota}. 
This helps improving background rejection 
capabilities. 

Considering liquid Argon (LAr) as the sensitive medium, the produced
light wavelength $\lambda$ is peaked at $\sim$128~nm. This presents technical
difficulties for PMTs, since conventional window glasses 
do not allow the passage of such a short wavelength.
To avoid this inconvenient, usually either the inner detector surface or the
PMT window is covered with a wavelength shifter, a substance able
to absorb and re-emit the light on a longer wavelength. One
of the most used wavelength shifters is the Tetraphenylbutadiene (TPB), which
re-emits the light at wavelengths around 420~nm~\cite{Burton:1973}.

Therefore, PMTs are a crucial element of noble-liquid detectors.
Even though the manufacturers provide data for each PMT model
(gain, dark current, linearity, etc.), the values presented in the
specification data sheet are usually too generic,
based on averages of many PMTs, not specific to the purchased one.
Non negligible deviations from the average behaviour can be expected
for a particular PMT. Moreover, companies do not usually provide
calibration data at cryogenic temperatures.
For these reasons they must be characterized individually as precisely as
possible in the laboratory.

Currently running noble liquid dark matter detectors
like WARP~\cite{WARP:2005} and XENON~\cite{XENON:2007}, employ small size
phototubes ($\sim2^{"}$ window) because of their relative small active volume (few liters). 
As a result of the scaling up of future detectors to improve 
dark matter detection
capabilities, the size of the installed photomultipliers will 
likely be increased:
the use of larger photocathode PMTs will allow to cover bigger surfaces
at lower cost. Indeed, the design of new prototypes like ArDM~\cite{Rubbia:2005ge} 
(one tonne of LAr) already incorporates the use of large area photocathode
tubes (8$^{"}$ window). The work presented in this paper is focused on the
study of those {\it large area} photomultipliers and how they behave when 
operated at cryogenic temperatures. 

In this paper, we first discuss the general features required for a PMT
as light sensor in a noble liquid dark matter detector
and present the nominal characteristics of the tested tubes
(section~\ref{sec:requirements}). 
The results of the main measurements performed on them are described in
section~\ref{sec:pmt_meas}:
quantum efficiency (\ref{sec:qe}),
single photoelectron spectrum (\ref{sec:ser}),
dark counts (\ref{sec:darkcounts})
and linearity (\ref{sec:linear}).
Conclusions are reported in section~\ref{sec:concl}.

% ==================================
\section{PMT requirements}
\label{sec:requirements}
% ==================================

In the framework of noble liquid dark matter experiments, the first property
required for a PMT is, obviously, to work properly at cryogenic
temperatures O(100~K). This strong requirement is generally not
satisfied by conventional PMTs.
Photocathodes are usually semiconductors, thus when temperature
is reduced to such low values, the cathode resistance can increase
by several orders of magnitude. This can cause a large voltage gradient
in the photocathode, resulting in a poor collection efficiency
at the first dynode~\cite{Spicer:1963, Ankowski:2006}. To avoid this undesirable
effect the photocathode has to be deposited on a conducting substrate.
Some companies are able to manufacture special coated PMTs for cryogenic
applications.

Another important feature one might require is a good PMT response
over a wide range of illumination levels. Signals ranging from the single
photoelectron (pe) up to several hundreds of photoelectrons 
(if the particle interaction takes place close to the photocathode)
can be expected in large detectors.
An adequate PMT gain ($G$) should allow the detection of both types of events,
the former without amplification and the latter without saturation.
The peak voltage of a signal can be estimated considering that the output charge
of the anode will follow a Gaussian distribution in time, with RMS equal to the transit time
spread of the electrons. If we assume a 50~$\Omega$ coupling
between the anode and the electronics, then the integral of the output signal will
be $50 \cdot G\cdot q$, with $q$ the electron charge.
The signal amplitude can be approximated by $A\sim 50 \cdot G \cdot q/TTS$
which for a $10^7$ gain and a typical transit time spread ($TTS$) of 5~ns
gives 16~mV output peak voltage. Hence, a nominal gain close to this value would
fulfill the requirements in the 1--100 photoelectrons range.
As explained later in the text, the gain is extracted from the
single photoelectron charge spectrum, the so called Single Electron Response (SER),
obtained with very low illumination levels.
Besides the gain, the single photoelectron charge spectrum allows the
study and characterization of other important PMT parameters
(peak to valley, peak spread, etc.) which determine, for instance, the
PMT energy resolution~\cite{Ostankov:2000}.

In order to estimate the number of photoelectrons produced in a single event,
the response of the PMT should also be proportional to the incident light.
Deviations from the ideal behavior are primarily caused by anode linearity
characteristics which, for pulsed sources it is mainly limited by space charge
effects due to the magnitude of the signal current. An intense light pulse 
increases the space charge density and causes current saturation.
Linearity should be at least granted up to levels of 100 pe.

Dark counts are signals that appear in the absence of light
and that are caused by
thermoionic emission~\cite{Hamamatsu}, 
leakage currents, glass envelope scintillation, 
field emission current, residual gases, and radioactivity
from the glass or the environment.
PMTs may be used as well for triggering and
although most of the dark counts will be suppressed
by time coincidence between several PMTs, a low rate is desirable.

Concerning the spectral response of the PMT, as mentioned previously, the
scintillation light emitted by excited Argon dimers has a wavelength peak
at $\sim$128~nm, where standard PMTs are blind.
Only tubes made of MgF$_2$ windows extend the 
range of visibility down to 110~nm.
Unfortunately, at the moment no large photocathode PMTs (8$^{"}$) are
manufactured with this special type of window, so typical experiments 
make use of coating materials like TPB to shift the light to the 400--450~nm
region where the PMT shows its best performance in terms of quantum 
efficiency (QE). This is a crucial parameter for dark matter experiments 
where very dim signals are recorded.

The timing resolution is another property which characterizes
the PMT performances but this parameter does not seem to be an issue.
Large photocathode PMTs actually present in the market
show $TTS$ values in the order of few ns, fast enough for
the requirements of a dark matter detector.

Finally, the phototubes must be placed inside the detector sensitive
volume so they should be manufactured using special low background materials.
Natural radioactivity from material compounds is a continuous source of low
energy neutron and alpha particles which can mimic true signal interactions.
Companies are able to manufacture a low background version
of many PMT models under request.

Among the inquired companies, only Electron Tubes Limited~\cite{ETL:2007}
and Hamamatsu~\cite{Hamamatsu:2007} can offer large photocathode PMTs suitable
to work under cryogenic conditions. Out of the available models,
we have chosen those closest to the required specifications:
the 9357-KFLB from Electron Tubes Limited (ETL) and
the R-5912-MOD from Hamamatsu.

Two units of each model have been tested.
Their main properties are summarized in table~\ref{tab:pmts}.
In particular, they have nominal gains around $10^7$ and peak quantum efficiencies
close to 20\%. Both models can be manufactured in low background glass. 

\begin{table}[ht]
\begin{center}
\begin{tabular}{c|c|c|c|c|c|c}
Manufacturer & Model & Size & Gain & TTS & Dark Currents & QE (@ 420nm)  \\
\hline\hline
ETL       & 9357 KFLB   & 8$^{"}$ & $1.1\times10^7$ & 4~ns  & 10 nA & 18\%  \\
\hline
Hamamatsu & R-5912-MOD & 8$^{"}$ &       $10^7$    & 2.4~ns & 50 nA & 22\% \\
\end{tabular}
\end{center}
\caption{PMT models tested on this work (values from manufacturer generic data sheet).}
\label{tab:pmts}
\end{table}

Hereafter, the tested ETL and Hamamatsu phototubes will be referred to as 
ETL1, ETL2 and HAM1, HAM2 respectively.

The voltage dividers have been made using a custom PCB double-sided printed
circuit board following the design provided by the manufacturer.
It is well known that the nominal values of those electronic components,
particularly the capacitance, might change substantially from room to cryogenic
temperature. Any change on these values has an impact on the voltage
applied to the dynode multiplication chain and hence on the PMT 
collection performance. In order to avoid this effect, specially manufactured
electronic components with guaranteed stability up to cryogenic temperatures
have been used.

% ==================================
\section{Measurements }
\label{sec:pmt_meas}
% ==================================

The need to measure all relevant properties for each particular
photomultiplier in the laboratory, beyond the generic specifications
provided by the manufacturer, is clear. Given our interest in 
cryogenic detectors, these measurements are mandatory since those 
specifications always refer to room temperature. 
The selected PMTs have been characterized to obtain their relevant
properties at both, room and cryogenic temperature
(liquid Argon bath temperature $\sim$87 K).
In the next sections, a detailed description of the experimental setup and
the results of the measurements is presented.

% ==================================
\subsection{Quantum Efficiency}
\label{sec:qe}
% ==================================

A precise knowledge of the PMT quantum efficiency as function
of the incident photon wavelength will be of capital importance for
a dark matter detector which aims to exploit the full capabilities
of a PMT, not only the trigger potentialities as a fast light detector
but also the calorimetric ones. Among other variables, the
parameterisation of QE as a function of $\lambda$ is needed 
to infer the total amount of scintillation light released 
inside the active detector volume, following a particle interaction.

The method used for measuring the spectral response of the PMT is
based on the comparison of its output signal with the response of a
calibrated detector, in our case a photodiode.
This measurement can be done by interchanging the calibrated and the unknown
photosensor after each measurement at a fixed value of $\lambda$,
or by splitting the light beam in two and taking both measurements simultaneously.
In order to minimize the errors induced by the light source instabilities and 
misalignments of the devices, the second method was followed.

As explained in section~\ref{sec:qe_lar}, the measurement is done,
first, at room temperature ({\it hot} spectral response) and then combined
with the data obtained in liquid Argon to compute the final value {\it in cold}.
As expected, it is found that the quantum efficiency properties
of the tested PMTs change substantially with temperature, so
a careful measurement of QE($\lambda$) in cryogenic
conditions is mandatory.

Sections~\ref{sec:qe_room} and~\ref{sec:qe_lar} describe the measurements
done at the cryogenic laboratory of the University of Granada
and the obtained results on the quantum efficiency measurements at room
and LAr temperatures, respectively.

% --------------------------------------------------
\subsubsection{Quantum Efficiency at room temperature}
\label{sec:qe_room}
% --------------------------------------------------

\begin{figure}[ht]
\leftmargin=2pc
\begin{center}
\includegraphics[width=14cm, clip]{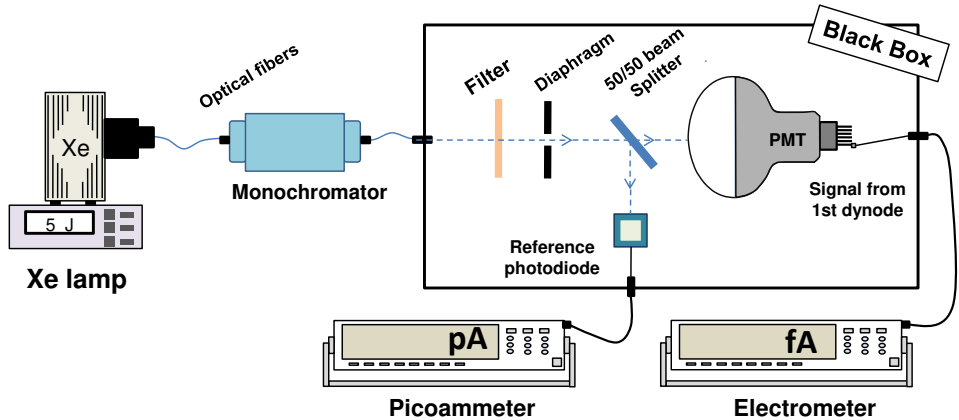}
\end{center}
\caption{Schematic drawing of the setup used for quantum efficiency
         measurements at room temperature.}
\label{fig:exsetuphot}
\end{figure}

Figure~\ref{fig:exsetuphot} shows the experimental setup used
to measure the quantum efficiency at room temperature.
It is mainly composed of a light source, a monochromator and a light-tight
box where the photosensors are placed.

The light source should provide an spectral distribution covering the
interesting range (300~nm--600~nm) with a high light intensity.
The ORIEL Newport 6427 Xenon lamp matches those requirements. It can
be used in pulsed mode (max. energy of 5~J) and the
output can be easily coupled to the monochromator by means of an
optical fiber. The stability of the lamp has been tested measuring
its output flux during several hours and it has shown deviations
below 1~\% in time slots of 5 minutes (duration of a single QE measurement).
The lamp is also equipped with a shutter.
 
The monochromator (Spectral Products Digikrom CM110 CVI),
controlled with a computer via the RS-232 port, is equipped with a
2400~g/mm grating which allows the selection
of a particular wavelength between 180 and 680~nm (peak at 240~nm)
with an accuracy of $\pm$0.6~nm.

A second optical fiber is used to send the monochromatic light
into the aluminum box which provides a completely dark environment
and acts as a Faraday cage. The box houses a filter (to block
any shorter {\it armonic} wavelength from the monochromator),
a diaphragm (to reduce the size of the light spot, such that it completely
fits inside the photodiode sensitive area), a 50/50 Polka-dot beam
splitter (by Edmund Optics) and the two photosensors. 
The reference detector is a Hamamatsu S1337-1010BQ photodiode, with
10$\times$10~mm$^2$ active area. It has been calibrated by the manufacturer.
The setup is completed with two devices to measure independently
the current produced by the reference photodiode and the photocathode
of the tested PMTs: a Picoammeter (mod. 6485) and a Electrometer (mod. 6514)
both from Keithley.

The quantum efficiency measurement comes from the comparison of the
currents produced in the photodiode and in the first dynode of the
unknown PMT. In order to ensure an optimal collection efficiency between
the photocathode ($K$) and the first dynode ($D_1$), 
a gradient of $\Delta V_{K-D_1}$= 300~V
is kept between these two points. Even in absence of light, this gradient
generates a small leak current ($\sim$10~pA) which must be subtracted from
the measurements in presence of light (shutter lamp open).

In order to get rid of systematic effects coming from the beam splitter
inaccuracy (5\% according to the manufacturer) two measurements were carried
out exchanging the positions of PMT and photodiode. The geometric
average of both measurements is taken as the final value. Indeed, in the
configuration shown in figure~\ref{fig:exsetuphot}, the photodiode receives
a fraction $\alpha$ of the incident light measuring an intensity $I^{r}$ 
(reflected in the splitter)
whereas the PMT receives the transmitted one (1-$\alpha$) and measures $I^{t}$.
 We can write:

\begin{equation}
  \label{eq:qetrans}
  \frac{I_{PMT}^{t}}{QE_{PMT}\cdot (1-\alpha)}=\frac{I_{PD}^{r}}{QE_{PD}\cdot\alpha}
\end{equation}
When the two photodetectors are exchanged:

\begin{equation}
  \label{eq:qerefl}
  \frac{I_{PMT}^{r}}{QE_{PMT}\cdot \alpha}=\frac{I_{PD}^{t}}{QE_{PD}\cdot(1-\alpha)}
\end{equation}
Combining equations~\ref{eq:qerefl} and~\ref{eq:qetrans} it follows:

\begin{equation}
  \label{eq:qehot}
  QE_{PMT}=QE_{PD}\cdot\sqrt{\frac{I_{PMT}^r}{I_{PD}^t}\frac{I_{PMT}^t}{I_{PD}^r}}
\end{equation}
which is independent of the beam splitter accuracy.

\begin{figure}[ht]
  \leftmargin=2pc
  \begin{center}
    \begin{tabular}{cc}
      \includegraphics[width=.5\textwidth]{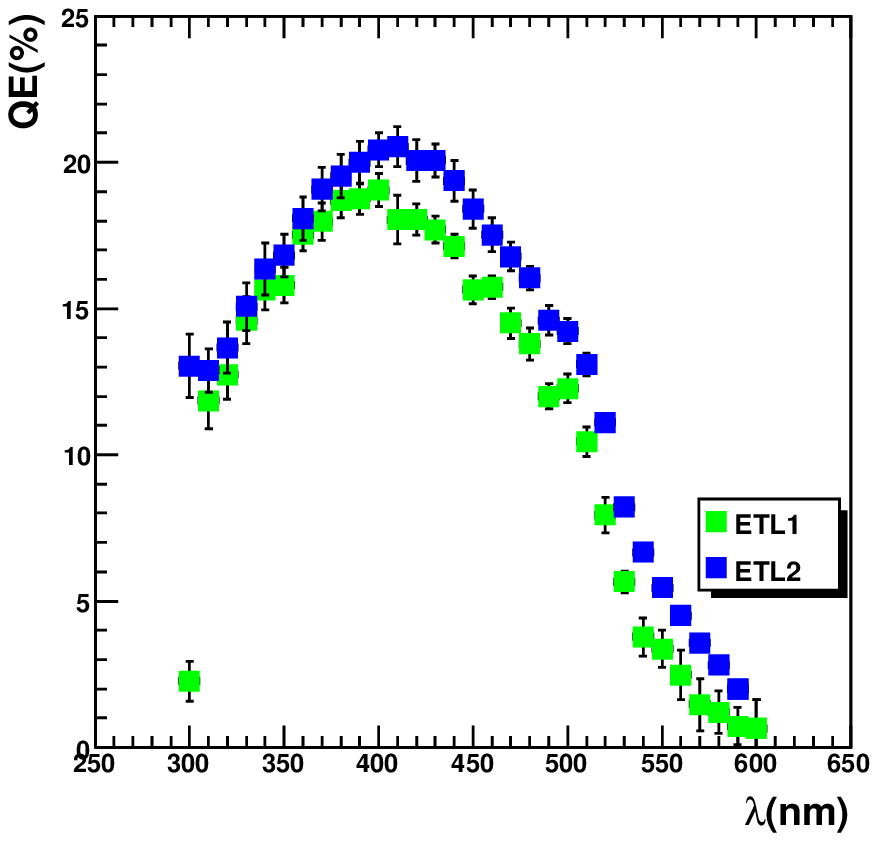}&
      \includegraphics[width=.5\textwidth]{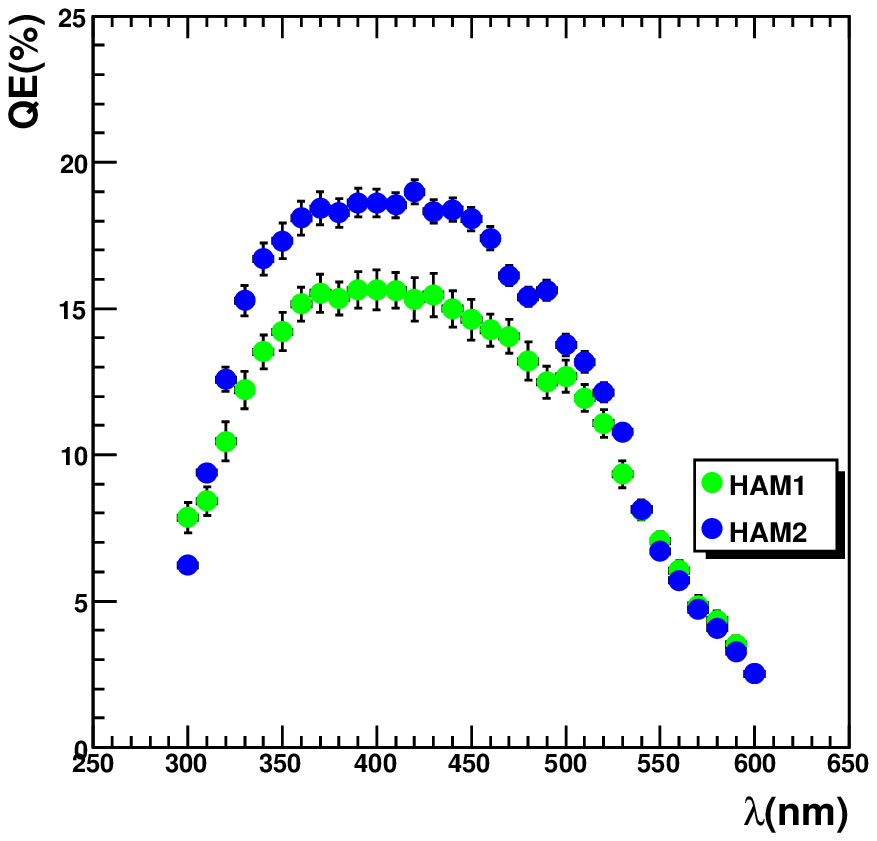}
    \end{tabular}
  \end{center}
  \caption{Quantum efficiency of the ETL (left) and Hamamatsu (right) PMTs
           as measured at room temperature.}
  \label{fig:qehot}
\end{figure}

Figure~\ref{fig:qehot} shows the results obtained for the measurements of quantum
efficiency at room temperature.

The points, in steps of 10~nm, follow a smooth curve. The quantum efficiency
shows maximum values in the range 15--20\% for a wavelength of about 400~nm.
In the vicinity of the peak, the Hamamatsu models show
an almost flat region between 350--450~nm, whereas the ETL PMTs present
a steeper behaviour.

% --------------------------------------------------
\subsubsection{Quantum efficiency at LAr temperature}
\label{sec:qe_lar}
% --------------------------------------------------

\begin{figure}[ht]
\leftmargin=2pc
\begin{center}
\includegraphics[width=14cm, clip]{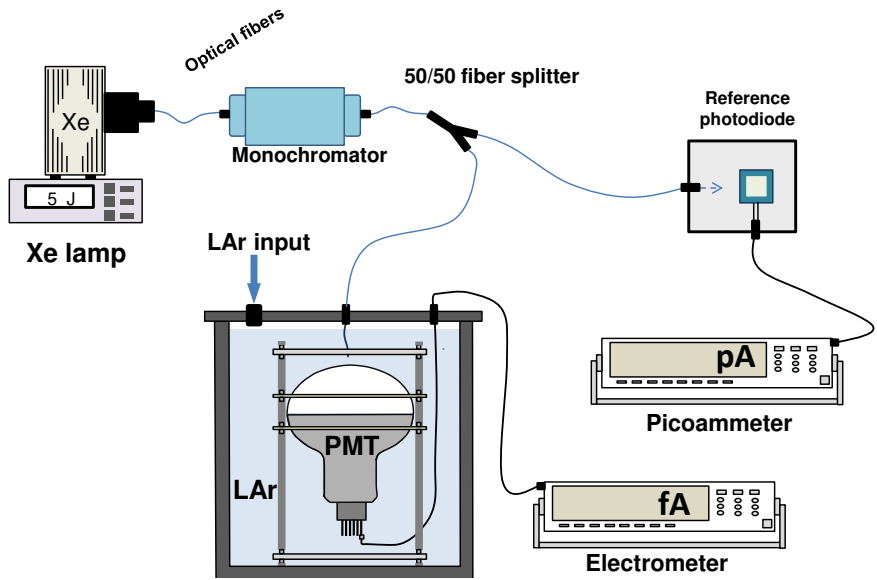}
\end{center}
\caption{Schematic drawing of the setup used for quantum efficiency
         measurements at LAr temperature.}
\label{fig:exsetupcold}
\end{figure}

To perform the measurements {\it in cold}, the previous experimental setup
has to be slightly modified (see figure~\ref{fig:exsetupcold}).
This measurement requires the tested PMT to be fully
immersed in liquid Argon for long periods of time (several hours).
A special light-blind cryostat is used for that purpose, ensuring stable
temperature conditions. It is equipped with three feedthroughs:
one for LAr filling,
a second one to feed the 1$^{st}$ dynode and read-out the current
with the Electrometer,
and the third one to pass the optical fiber which illuminates the PMT.
Inside the cryostat, the fiber is positioned very close to the PMT
surface such that dispersion effects due to 
a change of the medium conditions are negligible.
A specially designed supporting structure made of stainless steel
and polyethylene disks houses the PMT and fixes the fiber inside
the cryostat.

On the other hand, the Hamamatsu S1337-1010BQ reference photodiode
remains at room temperature in order to keep valid the manufacturer
calibration. It is placed inside an aluminum light-tight box.

The Xe lamp, the monochromator and the charge readout devices are
the same as in~\ref{sec:qe_room}. The Polka-dot beam splitter
is replaced by a 50/50 optical fiber splitter (Ocean Optics) allowing
a stable light flux between the monochromator and the two
photosensors.

A first measurement is done with the PMT in air, at room temperature,
where the intensity generated in the photodiode $I_{PD}^{hot}$ 
can be written as:

\begin{equation}
  \label{eq:Ipd}
  I_{PD}^{hot}=N_r\cdot \alpha\cdot QE_{PD}
\end{equation}

being $N_r$ the number of photons per unit of time before entering the fiber splitter,
and $\alpha$ the fraction of photons split to the photodiode. 
This is kept at room 
temperature during the whole procedure, therefore the labels $hot$ and $cold$ 
are just used to distinguish the two measurements. The intensity 
in the PMT, $I_{PMT}$, will be given by:

\begin{equation}
  \label{eq:Ipmthot}
  I_{PMT}^{hot}=N_r\cdot \left(1-\alpha\right)\cdot QE_{PMT}^{hot}
\end{equation}

Combining both equations:

\begin{equation}
  \label{eq:Ihot}
  \frac{I_{PMT}^{hot}}{\left(1-\alpha\right)\cdot QE_{PMT}^{hot}}=
  \frac{I_{PD}^{hot}}{\alpha\cdot QE_{PD}}
\end{equation}

Once the spectral response is measured at room temperature, the cryostat
is filled with LAr. The time needed for the PMT and the sustaining structure
to cool down and stop boiling is about 1 hour. After that, the set of
measurements is repeated and an analogous expression can be written:

\begin{equation}
  \label{eq:Icold}
  \frac{I_{PMT}^{cold}}{\left(1-\alpha\right)\cdot QE_{PMT}^{cold}}=
  \frac{I_{PD}^{cold}}{\alpha\cdot QE_{PD}}
\end{equation}

Assuming the fraction $\alpha$ and the photodiode quantum efficiency constant
during the whole procedure, the combination of equations~\ref{eq:Ihot}
and~\ref{eq:Icold} provides the following expression for the QE at LAr temperature:

\def\QEformula{%
  QE_{PMT}^{cold}=QE_{PMT}^{hot}\cdot \frac{I_{PD}^{hot}
       \cdot  I_{PMT}^{cold}}{I_{PD}^{cold}\cdot  I_{PMT}^{hot}}
    }
  \begin{equation}
    \label{eq:1}
    \QEformula
  \end{equation}

\begin{figure}[ht]
\leftmargin=2pc
\begin{center}
  \begin{tabular}{cc}
\includegraphics*[width=.5\textwidth]{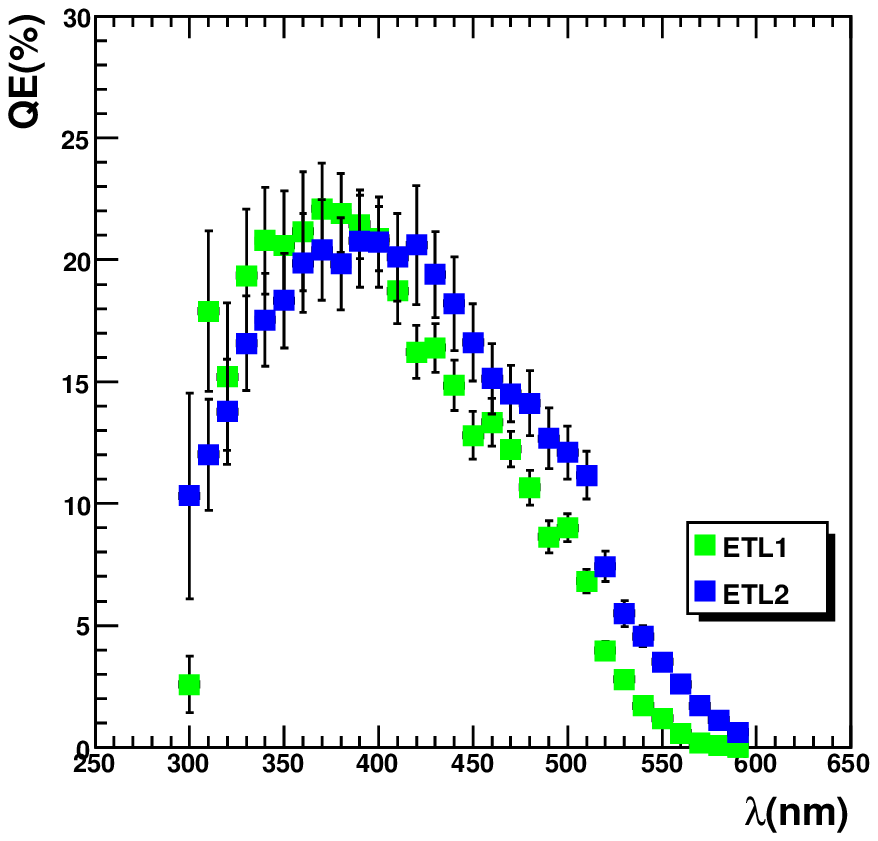}&
\includegraphics*[width=.5\textwidth]{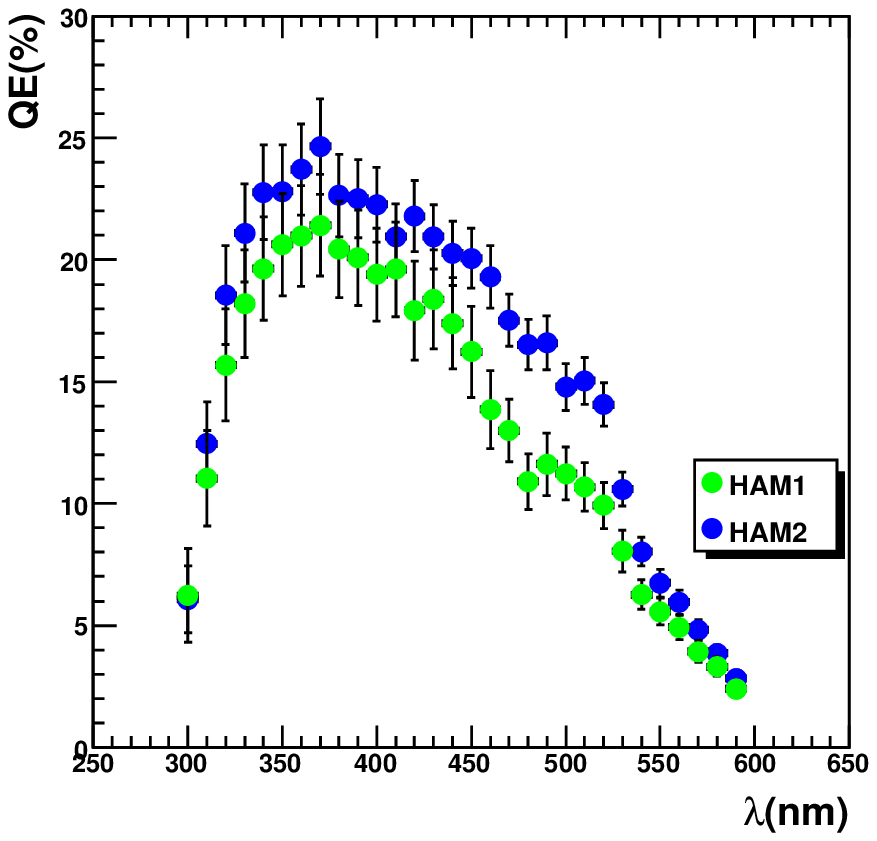}
  \end{tabular}
\end{center}
\caption{Quantum efficiency of the ETL (left) and Hamamatsu (right) PMTs
           as measured at cryogenic temperature.}
\label{fig:coldqe}
\end{figure}

Since measurements of QE$^{cold}$ are correlated to those at room 
temperature, errors have to be properly propagated. Figure~\ref{fig:coldqe} 
shows the QE results obtained for the tested PMTs.
For what concerns dark matter interactions, quantum efficiencies around
20$\%$ guarantee, for geometries like the one foreseen for the ArDM 
experiment, yields of about 1 photoelectron per keV of deposited energy.
This is enough to identify signals down to true recoil 
energies of 20 to 30 keV.  

\begin{figure}[ht]
\leftmargin=2pc
\begin{center}
  \begin{tabular}{cc}
\includegraphics*[width=.5\textwidth]{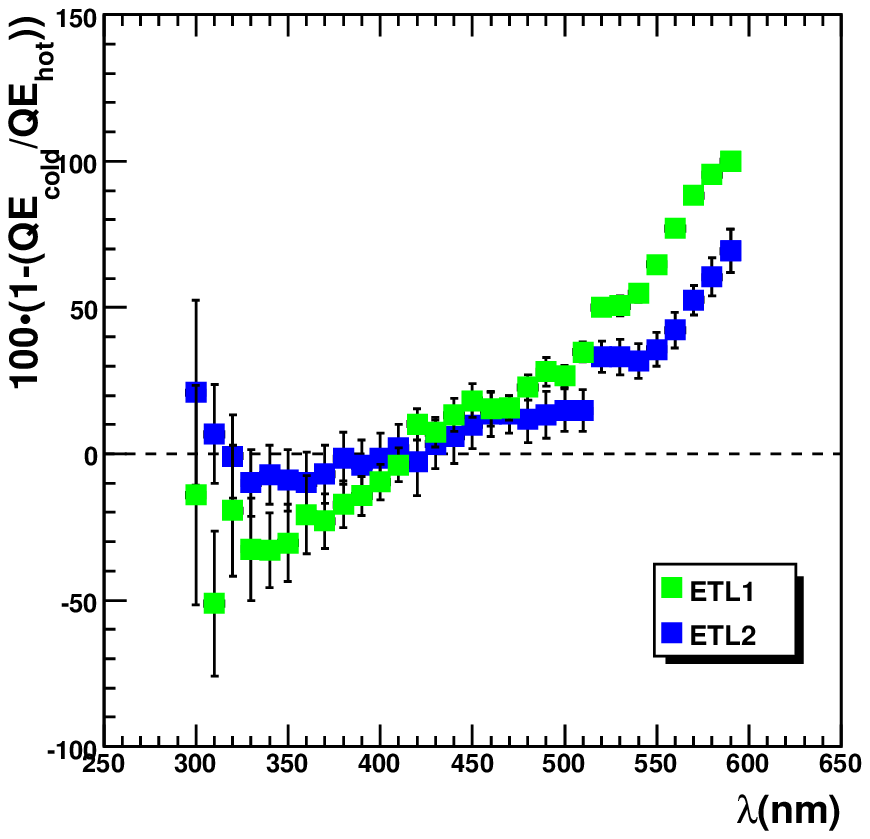}&
\includegraphics*[width=.5\textwidth]{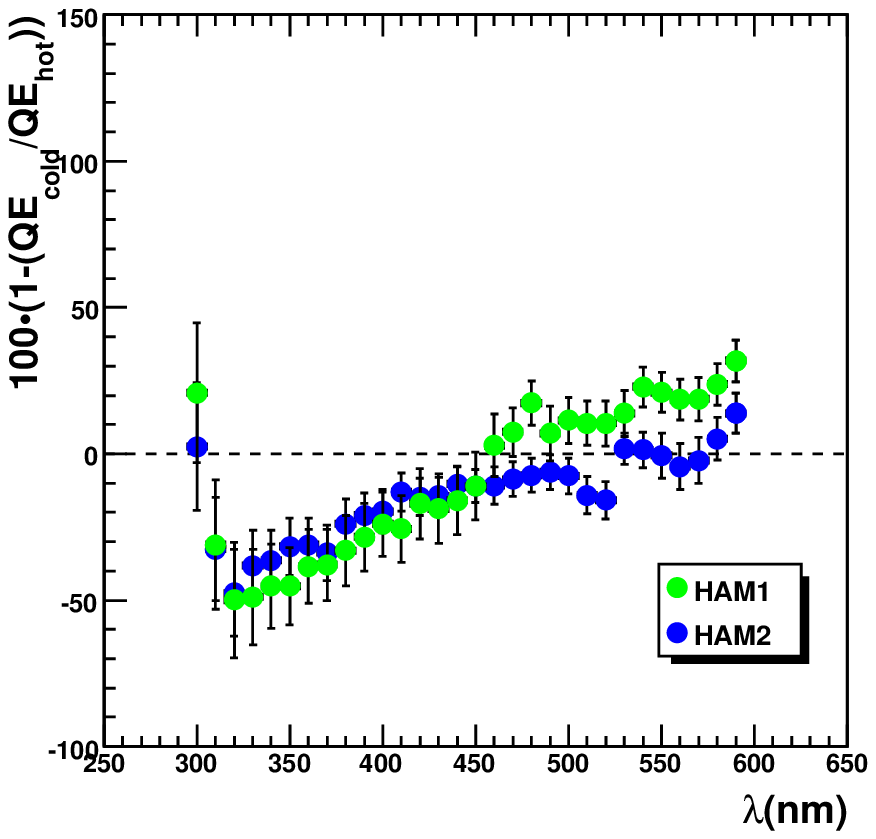}
  \end{tabular}
\end{center}
\caption{Change in quantum efficiency between room and cryogenic temperature
 for ETL (left) and Hamamatsu (right) models.}
\label{fig:coldchange}
\end{figure}

As compared to room temperature, a shift of the peak towards
shorter wavelengths and an increase of the absolute QE at the maximum
is observed in all cases. This effect is more clearly visible in
figure~\ref{fig:coldchange}, where the relative change from room to cryogenic
temperature $\left[ 100\cdot\left(1-\frac{QE_{cold}}{QE_{hot}}\right)\right]$ is shown.

This behaviour~\cite{Ankowski:2006, Araujo:1998, Singh:1987}
is related to the photoemission process~\cite{Spicer:1963} that 
takes place whenever a photon hits the photocathode of a PMT,
usually a semiconductor. When this happens, an electron is excited to the 
conduction band. The emission process is not a surface effect, but a bulk one. 
Hence, the electron must go through the semiconductor crystal until it reaches the vacuum, 
losing energy in each collision.
If the electron reaches the crystal surface with sufficient energy to escape 
over the potential barrier, it will be emitted from the photocathode.
When temperature decreases, the yield from the photocathode increases in the
region far from the cutoff wavelength, due to a decrease of the energy
loss of the electron in the lattice. In the region close to the cutoff 
wavelength, above which the photoelectric process is not anymore feasible, 
the yield decreases because of a decrease in occupied defect levels above
the valence band, an increase in band gap, 
and probably an unfavorable change in band bending.

% =================================================
\subsection{Response to single photoelectrons (SER)}
% =================================================
\label{sec:ser}

\subsubsection{Experimental setup}
% - - - - - - - - - - - - - - - - - - -

\begin{figure}[ht]
\leftmargin=2pc
\begin{center}
\includegraphics[width=\textwidth]{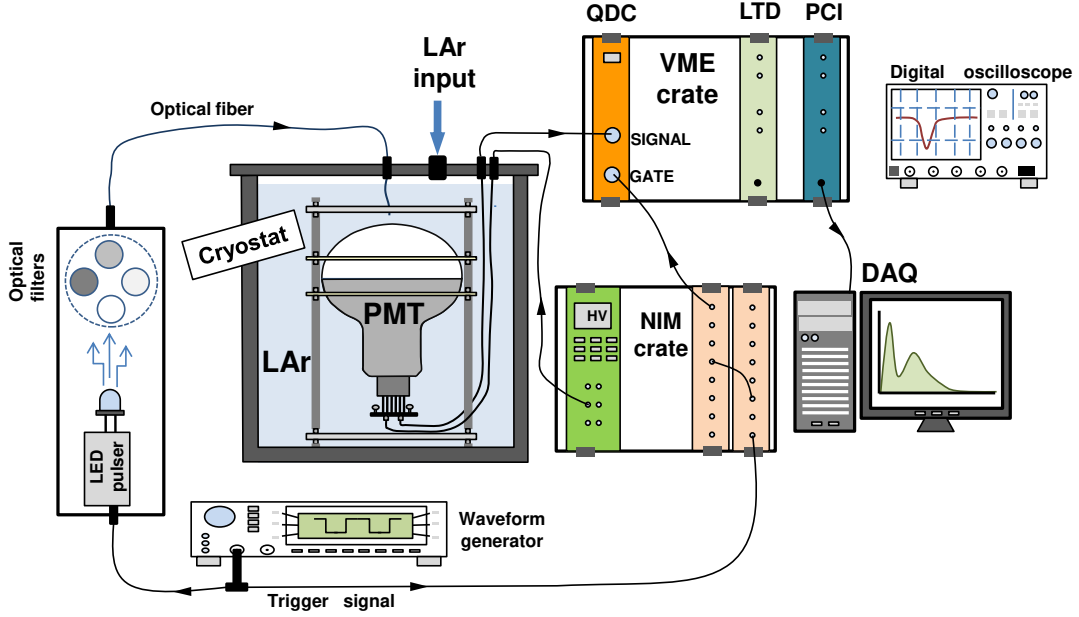}
\end{center}
\caption{Setup for gain, dark counts and linearity measurements.}
\label{fig:gainsetup}
\end{figure}

The setup used for single electron response, dark counts and linearity
measurements is shown in figure~\ref{fig:gainsetup}.
The PMT is housed inside a light-blind cryostat which can be filled
with liquid Argon. The measurements were sequentially done first at room
temperature and then in a LAr bath. The plots shown in this section
only refer to results at cryogenic temperature, although some comments on
comparisons with room temperature measurements are also included.

We use a blue LED as light source and connect it to a pulser which provides
1~kHz frequency, few nanoseconds width light shots.
If required, the light intensity can be attenuated by a set of neutral
density filters. The light is sent to the PMT photocathode inside the
cryostat thanks to an optical fiber. The PMT is polarized with a
NIM power supply (CAEN N470) and the signal transmitted through a standard
LEMO cable. Depending on the measurement to be done, the signal is
analyzed with a Charge-to-Digital Converter (QDC CAEN V965A),
an oscilloscope (LeCroy Waverunner 6100) or a
Low Threshold Discriminator (LTD CAEN V814B) plus a Scaler (CAEN V560AE).

\subsubsection{Gain}
% - - - - - - - - - - - - - - - - - - -

To measure the PMT gain, the amount of light reaching the photocathode
is reduced at the level of few photons. The number of generated 
photoelectrons follows a Poisson distribution and the probability 
of having $r$ photoelectrons is expressed as:

\begin{equation}
  \label{eq:poisson}
  P(r)=\mu ^r\cdot\frac{e^{-\mu}}{r!}
\end{equation}

The single photoelectron illumination level can be achieved by imposing
that most of the signals come from single electron events i.e.,
by requiring the number of signals with two photoelectrons being below,
for instance, 10\% of that of single photoelectron. Using
equation~\ref{eq:poisson}, the previous condition translates into $\mu=0.2$
and $P(0)=81.9\%$. Hence, if the light intensity is adjusted such that the number
of empty triggers is 81.9\% then the number of 2 photoelectrons signals will
be 10\% lower than that of 1 photoelectron. Under this condition, for any
other number of photoelectrons the probability will be negligible.

\begin{figure}[ht]
\begin{center}
\includegraphics[width=.5\textwidth]{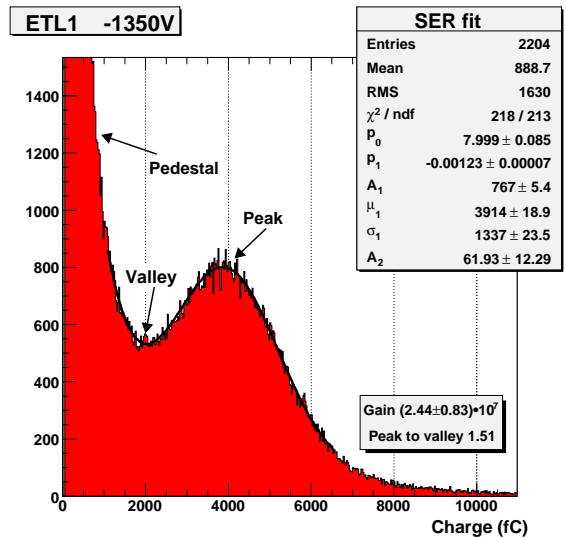}
\end{center}
\caption{Example of a SER spectrum as obtained for the ETL1 at 1350 V.}
\label{fig:ser}
\end{figure}

Figure~\ref{fig:ser} shows an example of the collected charge spectrum
under these conditions (SER)
as obtained for ETL1 at 1350 V.
The SER distribution has been fitted to a function which contains
the following terms~\cite{Ankowski:2006, Dossi:2000}:

\begin{itemize}
\item An exponential distribution that fits the {\it pedestal} and which is
caused by several factors: the continuous component of the dark current,
the intrinsic shift of the QDC, the electronic noise affecting the measurement, etc.
This exponential is parameterized as $e^{p_0+p_1\cdot x}$.

\item A Gaussian distribution which takes into account the response
of the PMT to a single photoelectron (parameters $A_1$, $\mu_1$ and $\sigma_1$).

\item An extra Gaussian with parameters $A_2, \mu_2=2\cdot\mu_1$ and
$\sigma_2=\sqrt{2}\cdot\sigma_1$ to account for events with 2 photoelectrons. 
In general, a sum of $n$ Gaussians can be included to reproduce
$n$-photoelectron contributions.

\end{itemize}

\begin{figure}[ht]
  \begin{center}
    \begin{tabular}{cc}
      \includegraphics[width=.5\textwidth]{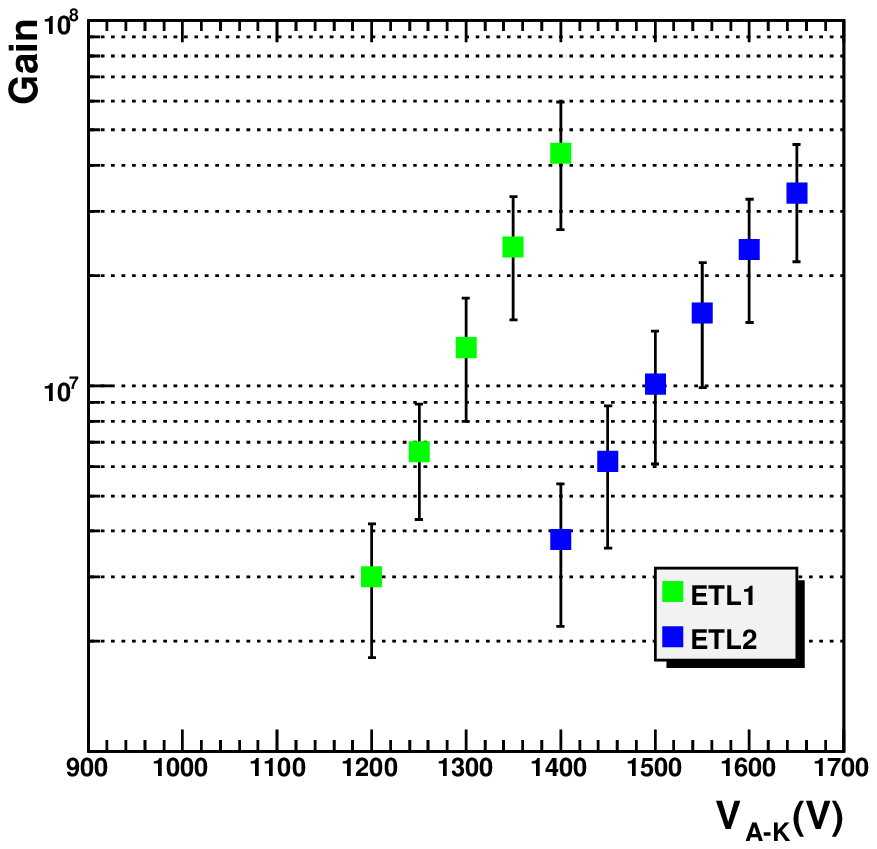}&
      \includegraphics[width=.5\textwidth]{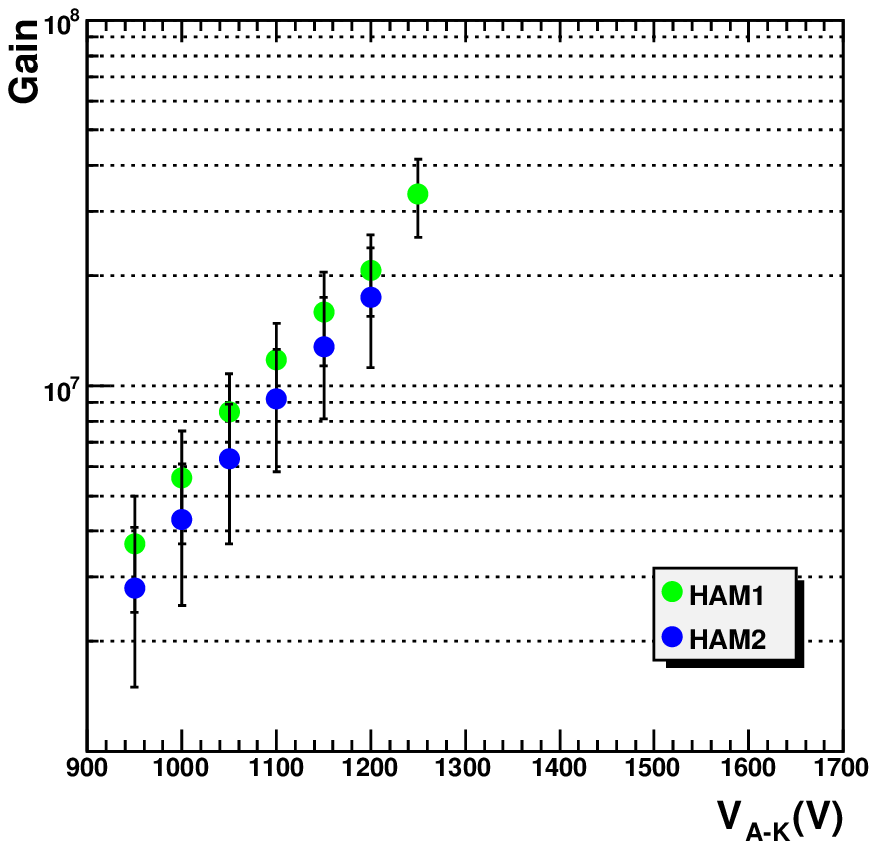}
    \end{tabular}
    \caption{Gain dependence on HV for the four photomultipliers as measured in LAr.}
    \label{fig:gain}
  \end{center}
\end{figure}

The gain is obtained from the position of the single photoelectron peak
in the charge spectrum. Figure~\ref{fig:gain} shows the gain dependence
on HV for the four tested PMTs in cold. A clean linear behaviour
is observed in all cases.
The Hamamatsu tubes achieve the nominal $10^7$ gain at about 1100~V
whereas the ETL tubes need higher voltages. On the other hand,
the slope for ETL PMTs is steeper. It is interesting to notice that similar
values and slopes were quoted at room temperature. This uniformity
on the results can be attributed to the stability of the cryogenic electronic
components mounted on the PMT base.

\subsubsection{Peak to Valley ratio and SER resolution}
% - - - - - - - - - - - - - - - - - - - - - - - - - - - -

\begin{figure}[ht]
  \leftmargin=2pc
  \begin{center}
    \begin{tabular}{cc}
      \includegraphics[width=.5\textwidth]{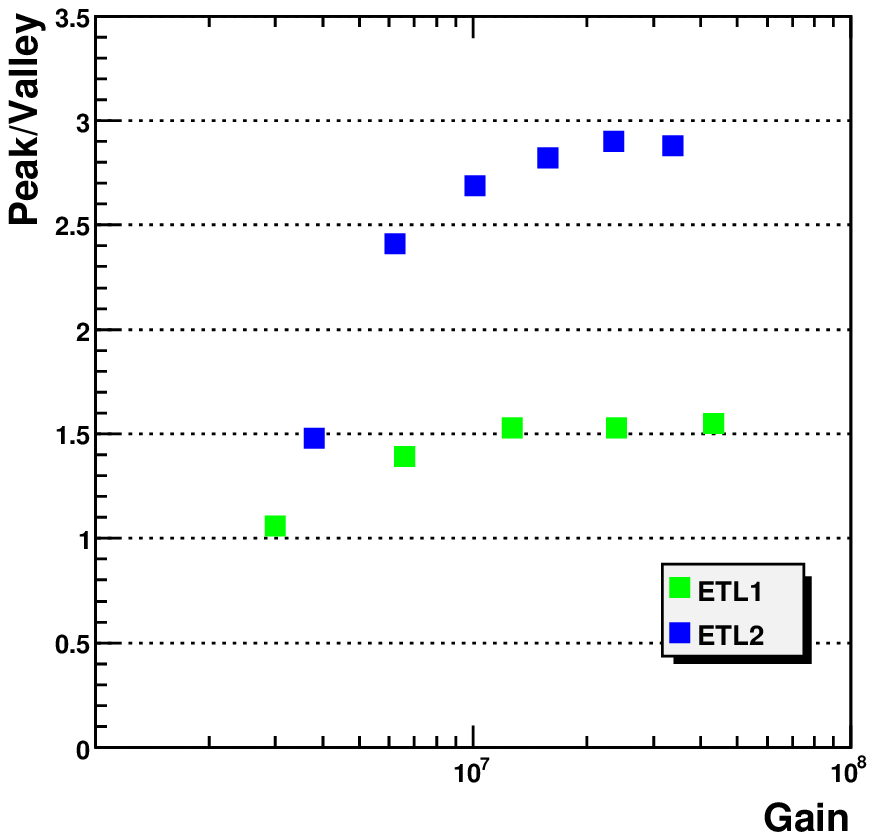}&
      \includegraphics[width=.5\textwidth]{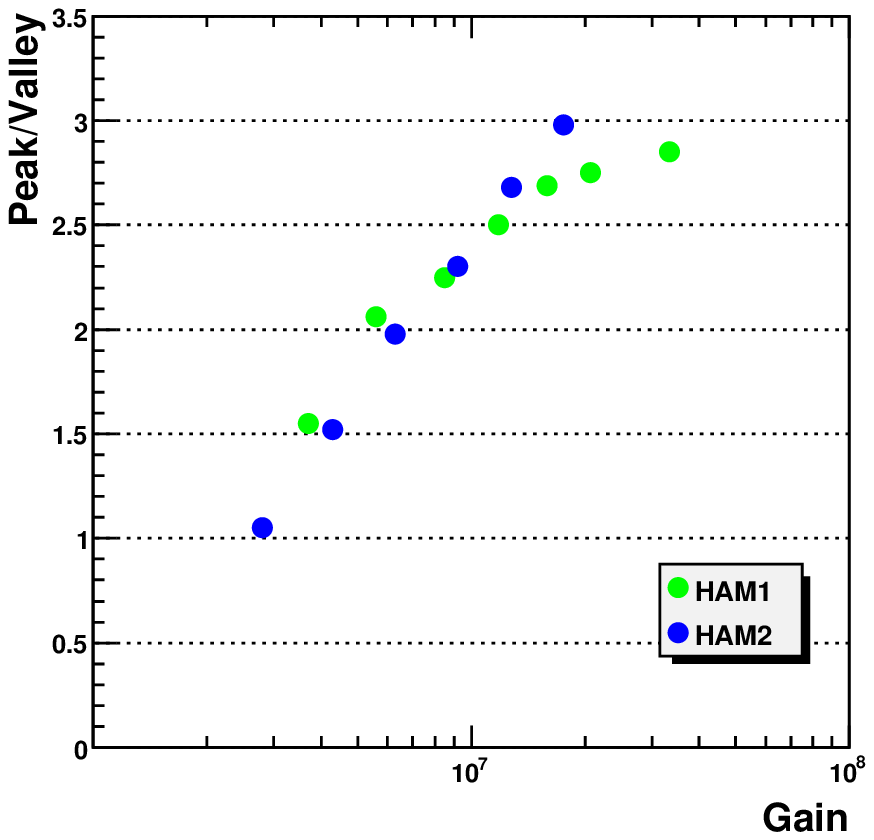}
    \end{tabular}
    \caption{Peak to valley dependence on gain at LAr temperature.}
    \label{fig:p2v}
  \end{center}
\end{figure}

The ratio between the height of the single photoelectron peak and the valley
(P/V), obtained from the charge spectrum (see figure~\ref{fig:ser}) is another
important characteristic to be measured. Figure~\ref{fig:p2v} shows the 
results obtained on this quantity as a function of the gain.

The peak to valley ratio increases with gain in all cases.
For a gain of $10^7$ the measured value is around 2.5 for all PMTs
but the ETL1, which shows a lower value. Beyond this point, the ETL
tubes show an almost flat P/V while the Hamamatsu ones increase with voltage.

\begin{figure}[ht]
  \begin{center}
    \begin{tabular}{cc}
      \includegraphics[width=.5\textwidth]{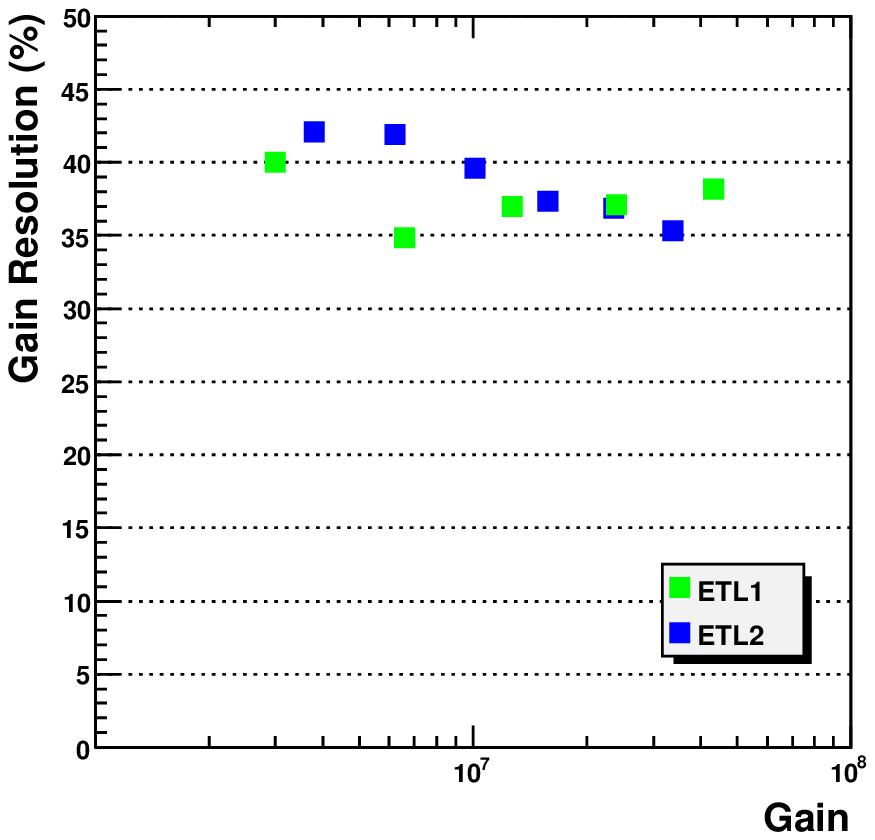}&
      \includegraphics[width=.5\textwidth]{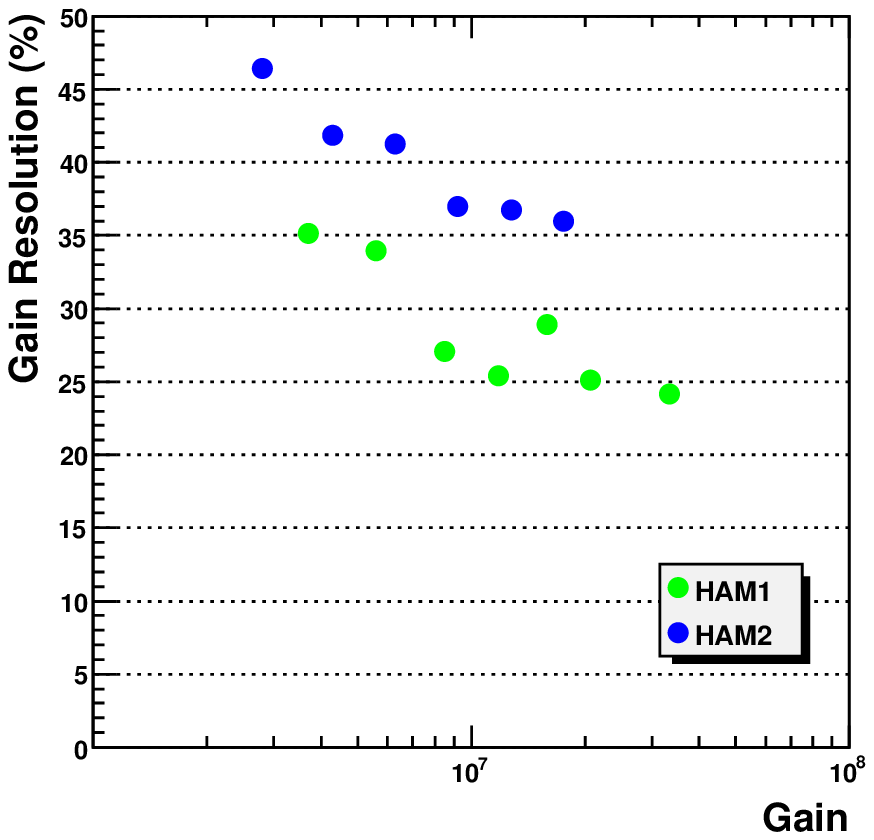}
    \end{tabular}
    \caption{Gain resolution from SER spectrum for the tested PMTs at LAr temperature.}
    \label{fig:resol}
  \end{center}
\end{figure}

Figure~\ref{fig:resol} shows the resolution in the SER peak, defined as 
the ratio $\frac{\sigma_1}{\mu_1}$ in percentage.
While the ETL tubes exhibit an almost flat dependence on the considered range,
the resolution decreases smoothly on the Hamamatsu PMTs as gain is increased.
At the nominal value of $10^7$ the four tubes show resolution values in the range
35--40\%.

% ==================================
\subsection{Dark Counts}
% ==================================
\label{sec:darkcounts}

\begin{figure}[ht]
\leftmargin=2pc
\begin{center}
\includegraphics[width=.5\textwidth]{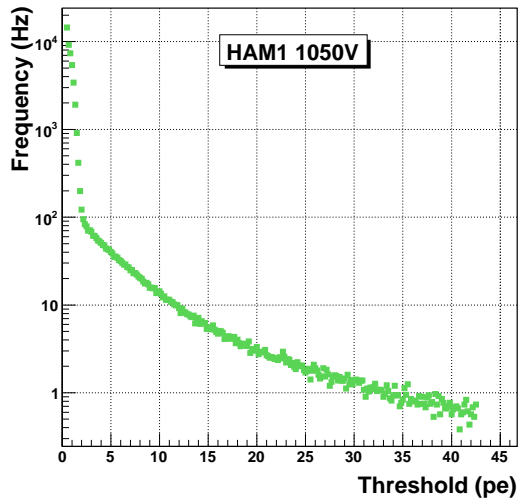}
\end{center}
\caption{Dark counts rate dependence on discriminator threshold for HAM1 at 1050~V.}
\label{fig:dc_single}
\end{figure}

Prior to measure the dark count frequency, the phototubes are placed
in darkness, inside the container filled with LAr and polarized for several hours.
The PMT output is then connected to the discriminator and the output
signal is feed to a scaler (CAEN V560AE), where the number of pulses above a given
threshold is counted. As an example, figure~\ref{fig:dc_single} shows the
result for HAM1 polarized at 1050~V.

To compare different PMTs at different voltages, thresholds in the
discriminator (mV) have to be expressed in terms of number of photoelectrons
in amplitude. The SER peak voltage is obtained from recorded oscilloscope signals
taken at very low intensity illumination conditions.
Looking at figure~\ref{fig:dc_single}, the abrupt fall (factor 100) at low thresholds, 
corresponding to the level of single photoelectron, is clear. Beyond this point, the
decrease in rate when increasing the threshold is smooth.

\begin{figure}[ht]
  \leftmargin=2pc
  \begin{center}
    \begin{tabular}{cc}
      \includegraphics[width=.5\textwidth]{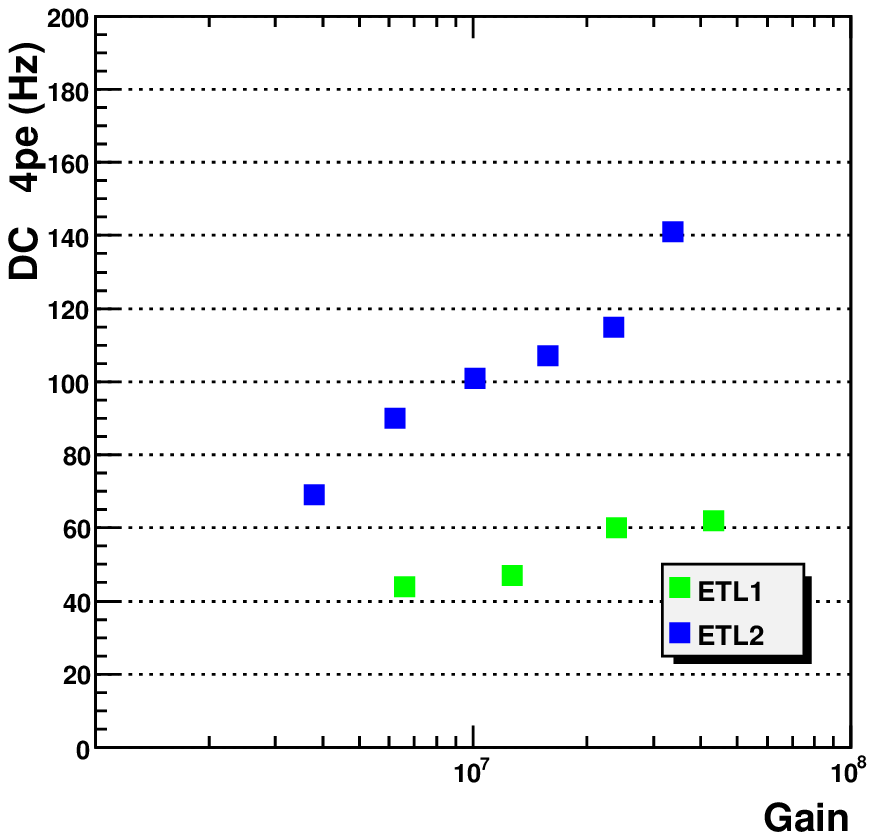}&
      \includegraphics[width=.5\textwidth]{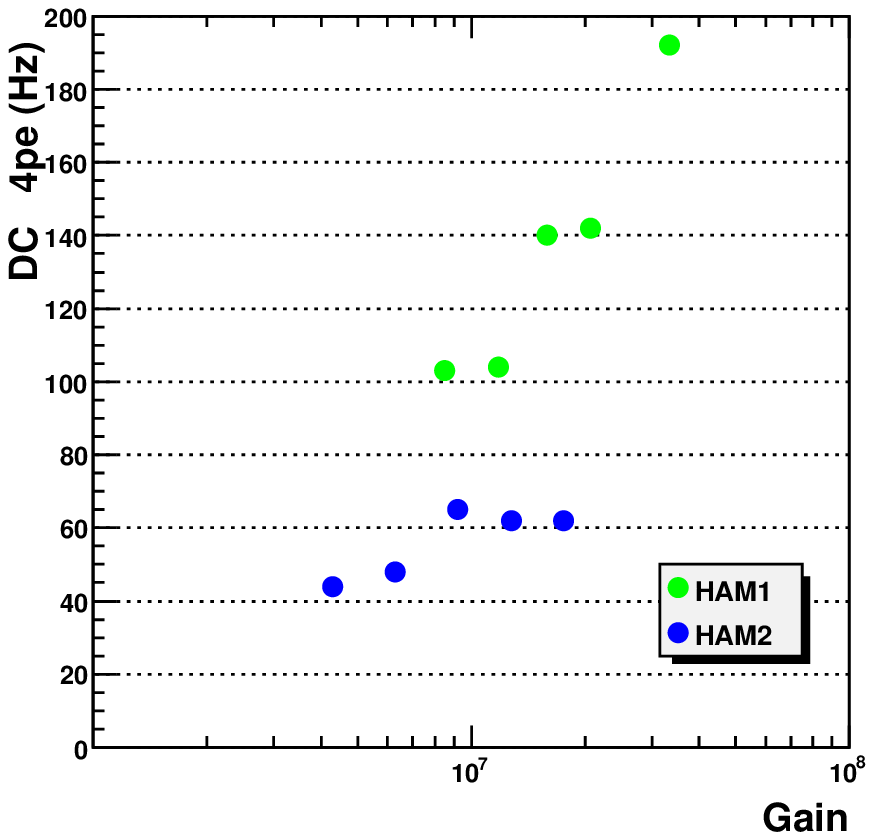}
    \end{tabular}
    \caption{Dark counts rates as measured in LAr (4 pe threshold).}
    \label{fig:dc4}
  \end{center}
\end{figure}

Figure~\ref{fig:dc4} shows the number of dark counts for different values of gain
above 4 photoelectrons for every PMT. This threshold has been selected as it is far
from the abrupt fall region but it is still low enough to be used as a trigger.

As expected, the dark count rates increase with gain (higher voltages).
Frequencies in the range 50--100~Hz are obtained for gain values around $10^7$.

Compared with the results at room temperature a clear decrease in the rates
is measured {\it in cold}: a factor close to 5 and 2 for the ETL and
Hamamatsu tubes, respectively. This effect is explained by the decrease
of the thermal energy of the electrons in the photocathode.

% ==================================
\subsection{Linearity}
% ==================================
\label{sec:linear}

The study of the PMT response to different illumination levels
above the single photoelectron was carried out using neutral
density filters~\cite{Dossi:2000}.
They were placed in a rotating support, just between the light source
(pulsed blue LED) and the optical fibre light guide
(see figure~\ref{fig:gainsetup}). Five filters were used
(optical density values {\it d} = 3.0, 2.0, 1.5, 0.5 and 0.3)
which allowed variations of the light intensity brought to the photocathode
by three orders of magnitude (attenuation factor = $10^{d}$).
The measurement proceeds as follows: the PMT voltage is adjusted to a gain
of $10^7$ and maintained unchanged, the higher optical density filter selected
and the LED intensity tuned and fixed. In these conditions 
the average number of photoelectrons in the PMT is fixed between one and two. 
Then, a PMT charge spectrum is recorded with each filter and analyzed.

\begin{figure}[ht]
\leftmargin=2pc
\begin{center}
\includegraphics[width=.5\textwidth]{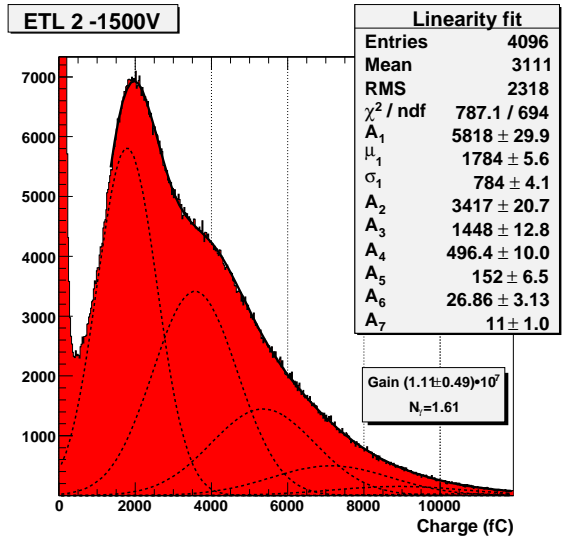}
\end{center}
\caption{Example of charge spectrum (ETL2) obtained, during the 
linearity measurements, with the 3.0 optical density filter. 
Dotted lines (solid line) correspond to individual Gaussian fits (global fit).} 
\label{fig:lin30}
\end{figure}

Figure~\ref{fig:lin30} shows the kind of spectrum obtained by this method.
If the parameters of the Gaussian for one photoelectron are $\mu_1$ and $\sigma_1$,
the parameters of the distribution for the coincidence of $n$ photoelectrons are
given by the sum of the corresponding Gaussians, i.e. 
$\mu_n=n\cdot\mu_1$ and $\sigma_n=\sqrt{n}\cdot\sigma_1$.
The average number of photons can be then computed
as $\frac{\Sigma i\cdot N(i)}{\mu_1 \cdot \Sigma N(i)}$, where $N(i)$ is the
content of the {\it i}-th bin of the charge distribution.
For the measurement shown in figure~\ref{fig:lin30}, a value close to 1.6 is obtained.
Dotted lines correspond to the first five Gaussians
whereas the convolution (solid line) nicely follows the global data.
The high peak on the most left corresponds to the pedestal (empty trigger events).

In order to complete the linearity measurements, charge spectra with
all neutral density filters are acquired.
For very high photocathode illumination levels it is not possible anymore
to resolve peaks for different number of photoelectrons and there are no
empty trigger events, which requires a prior measurement of the pedestal.
The ratio between the average charge in the distribution and the charge
of a single photoelectron (from the $d$ = 3.0 filter measurement)
gives the mean number of photoelectrons.

\begin{figure}[ht]
    \begin{tabular}{lr}
      \includegraphics[width=.5\textwidth]{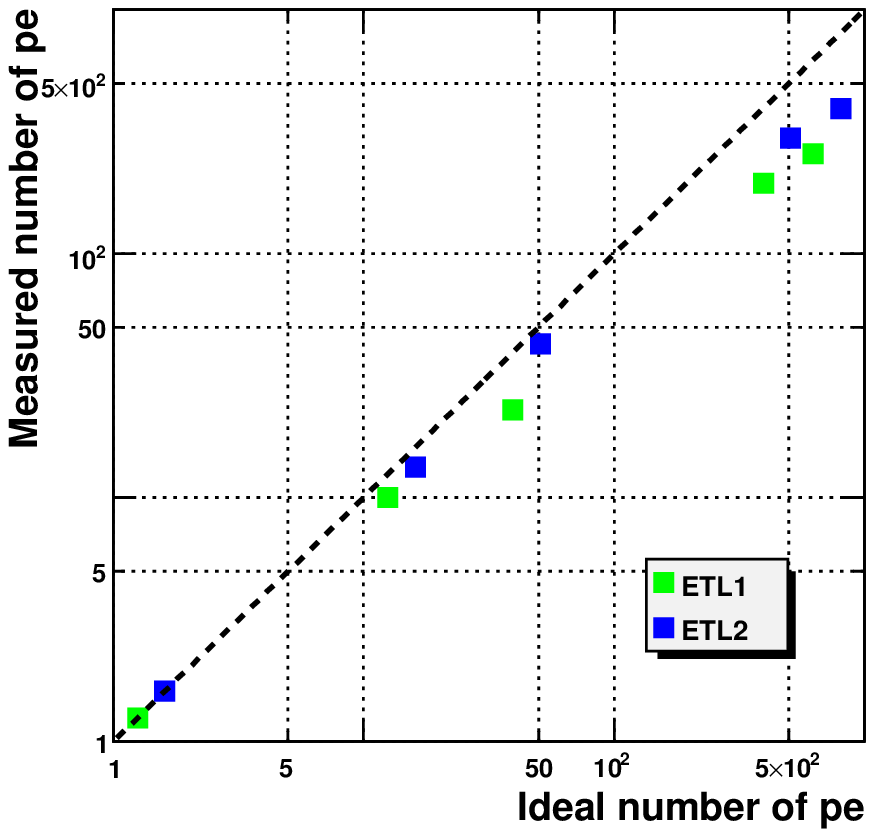}&
      \includegraphics[width=.5\textwidth]{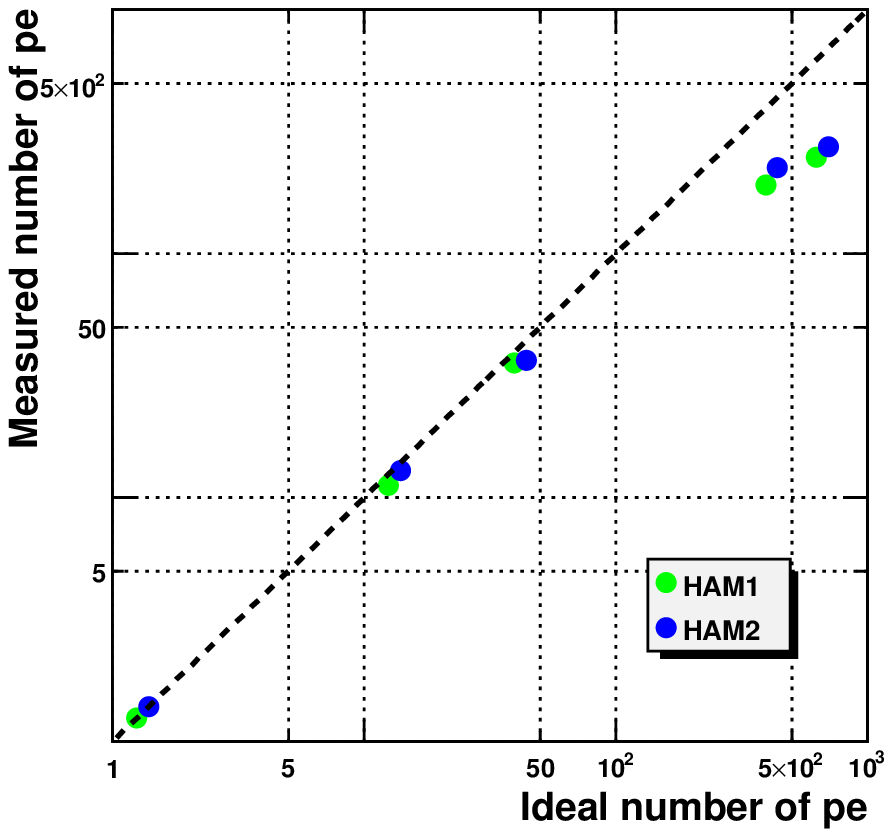}
    \end{tabular}
    \caption{Linearity measurements: PMT signal in units of number
     of photoelectrons as function of the ideal one as measured in liquid argon.
     The dotted line shows the perfect linearity behaviour.}
    \label{fig:linearity}
\end{figure}

A comparison between the ideal number of photoelectrons and the measured one
can be obtained using the transmittance relationship between the filters.
The result is shown in figure~\ref{fig:linearity}.
The four PMTs show a nice linear behaviour up to 100 pe. Above this 
value a slight departure from linearity is clearly seen. However this is 
not a serious concern, since such large energy depositions are well 
above the energy window where we expect to see most of the signal due 
to interactions of dark matter components. 

% ===============================
\section{Conclusions}
\label{sec:concl}
% ===============================

Current dark matter experiments based on noble liquids 
rely exclusively on photomultipliers to readout light signals. 
An improvement of several orders of magnitude on limits 
currently provided by experiments directly searching 
for dark matter signatures,
require to increase detectors masses up to the tonne scale or bigger. 
Accordingly, large-area photomultipliers are an optimal and 
cost-effective solution to detect scintillation light. 

In this paper we have analyzed whether large-area phototubes are 
suitable for operation at cryogenic temperatures and fulfil the stringent 
physics requirements imposed by future dark matter experiments. Only 
two companies are able, nowadays, to provide PMTs capable of operating 
at temperatures in the vicinity of hundred kelvins or lower. We 
have extensively tested 8$^{"}$ phototubes from ETL (9357 KFLB) 
and Hamamatsu (R-5912-MOD) both at room and liquid Argon temperatures. 
Our conclusion is that both models are adequate for installation in 
an experiment like ArDM, since they have QE above 20$\%$ for 400 nm, 
thus guaranteeing light yields of at least 1 pe/keV, enough to detect 
signals depositing around 20 to 30 keV. They show a linear behaviour 
throughout the energy window where we expect to have better chances 
to detect dark matter. In addition, their timing accuracy is such that 
they will allow to distinguish fast from slow scintillation light components, 
thus providing a powerful rejection tool against background.

Finally, we summarize the most relevant phototube 
features found at cryogenic temperatures:

\begin{itemize}

\item An increase of the maximum values of the quantum efficiencies
up to 25\% together with a global
shift of the distributions towards shorter wavelengths is observed as
compared with results at room temperature. ETL models appear to have
higher peak values.

\item A gain of $10^7$ is reached for both models with polarization
voltages between 1100 and 1500~V, being the lowest values for the Hamamatsu PMTs.

\item The dark count rate at 4 photoelectron threshold has a frequency
of about 100~Hz, similar for both models.

\item A good linear behaviour is achieved even at very high illumination levels,
up to $\sim$100 photoelectrons.
\end{itemize}

% ===============================
\section{Acknowledgements}
\label{sec:acknowledgments}
% ===============================

This work has been supported by CICYT Grants FPA-2006-00684,
FPA-2002-01835 and FPA-2005-07605-C02-01. We warmly thank our 
colleagues from the ArDM experiment for useful discussions and advise 
during the realization of these tests. 

% ===============================

\end{document}